\documentclass[aip,amsmath,amssymb,twocolumn,]{revtex4}                % two column

\usepackage{textcase}
\usepackage{graphicx}% Include figure files
\usepackage{dcolumn}% Align table columns on decimal point
\usepackage{bm}% bold math
\usepackage{amsmath,amsthm,amssymb,mathrsfs}

\begin{document}

\title{Linewidth of Power Spectrum Originated from Thermal Noise in Spin Torque Oscillator}

\author{Tomohiro Taniguchi}

\affiliation{ 
National Institute of Advanced Industrial Science and Technology (AIST), Spintronics Research Center, Tsukuba 305-8568, Japan}

\date{\today}% 

\begin{abstract}
A theoretical formula of the linewidth caused by the thermal activation in a spin torque oscillator with a perpendicularly magnetized free layer 
and an in-plane magnetized pinned layer was developed by solving the stochastic Landau-Lifshitz-Gilbert equation in the energy-phase representation. 
It is shown that the linewidth can be suppressed down to 0.1 MHz by applying a large current (10 mA for typical material parameters). 
A quality factor larger than $10^{4}$ is predicted in the large current limit, which is two orders of magnitude larger than the recently observed experimental value. 
\end{abstract}

\maketitle

% ===================================================================================================================================================================================== %

The spin torque oscillator (STO) \cite{kiselev03,rippard04,houssameddine07,deac08,kudoPRB10,sinha11} 
is a promising candidate for a future nanocommunication device 
because of its small size, high emission power, and frequency tunability. 
Recently, it was found \cite{kubota13,tamaru13MMM} that 
an STO with a perpendicularly magnetized free layer 
and an in-plane magnetized pinned layer \cite{yakata09,ikeda10,kubota12,zeng13,taniguchi13,taniguchi14IEEE} 
can achieve a large emission power of close to 1 $\mu$W. 
Therefore, this type of STO 
will be the model structure for practical STO applications. 

% ===================================================================================================================================================================================== %

Another important quantity characterizing the STO's properties 
is the linewidth $\Delta f$ of the power spectrum. 
For example, a narrow linewidth is necessary to obtain 
a high quality factor (Q-factor) $Q=f_{0}/\Delta f$, 
where $f_{0}$ is the peak frequency of the power spectrum. 
The physical origin of the linewidth is the nonuniform magnetization dynamics 
due to thermal activation. 
Therefore, theoretical evaluation of the linewidth due to thermal activation 
and of the Q-factor is desirable 
to clarify the theoretically possible values of these parameters. 

% ===================================================================================================================================================================================== %

In the self-oscillation state of an STO, 
the energy supplied by the spin torque balances 
the energy dissipation due to damping; 
therefore, the magnetization steadily precesses almost on a constant energy curve. 
This situation is very similar to magnetization switching in the thermally activated region, 
where the magnetization precesses on a constant energy curve many times 
during switching. 
Recent studies \cite{apalkov05,bertotti06,dykman12,taniguchi13PRB1,taniguchi13PRB2,taniguchi14JAP} 
have shown that the energy-phase representation of the Landau-Lifshitz-Gilbert (LLG) equation is useful 
for theoretical investigations of the magnetization switching properties, such as the switching probability, 
in the thermally activated region. 
Accordingly, we were motivated to develop a theory of the linewidth of an STO 
based on the LLG equation in the energy-phase representation. 

% ===================================================================================================================================================================================== %

In this letter, 
the theoretical formula for the linewidth of an STO is derived 
on the basis of the LLG equation in the energy-phase representation. 
It is shown that the linewidth can be suppressed to 0.1 MHz 
by applying a large current ($\sim$ 10 mA). 
The Q-factor can reach more than $10^{4}$ in this type of STO, 
which is two orders of magnitude larger than the previously reported value \cite{kubota13}. 

% ===================================================================================================================================================================================== %

\begin{figure}%[p]
\centerline{\includegraphics[width=1.0\columnwidth]{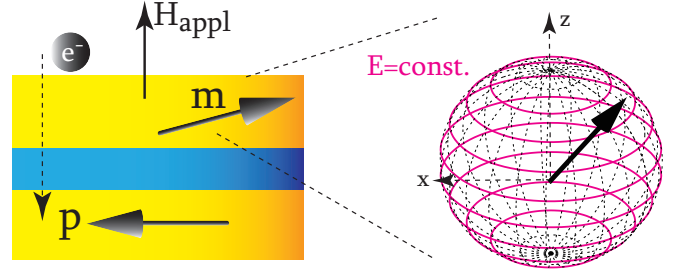}}\vspace{-3.0ex}
\caption{
         Schematic views of STO under consideration 
         and unit sphere on which the magnetization moves. 
         Solid lines in the unit sphere represent the constant energy curves. 
         \vspace{-3ex}}
\label{fig:fig1}
\end{figure}

% ===================================================================================================================================================================================== %

% ===================================================================================================================================================================================== %

Figure \ref{fig:fig1} shows a schematic view of the STO under consideration, 
where the unit vectors pointing in the magnetization directions 
of the free and pinned layers are denoted as $\mathbf{m}$ and $\mathbf{p}$, respectively. 
The $z$-axis is normal to the film-plane, 
and the $x$-axis is parallel to $\mathbf{p}$. 
The external field $H_{\rm appl}$ is applied along the $z$-axis. 
The current $I$ flows uniformly along the $z$-axis, 
where positive current corresponds to 
the electron flow from the free layer to the pinned layer. 
The LLG equation is given by 
\begin{equation}
\begin{split}
  \frac{d \mathbf{m}}{d t}
  =&
  -\gamma
  \mathbf{m}
  \times
  \mathbf{H}
  -
  \gamma
  H_{\rm s}
  \mathbf{m}
  \times
  \left(
    \mathbf{p}
    \times
    \mathbf{m}
  \right)
\\
  &-
  \gamma
  \mathbf{m}
  \times
  \mathbf{h}
  +
  \alpha
  \mathbf{m}
  \times
  \frac{d \mathbf{m}}{dt},
  \label{eq:LLG}
\end{split}
\end{equation}
where $\gamma$ and $\alpha$ are 
the gyromagnetic ratio and Gilbert damping constant, respectively. 
The magnetic field is defined as 
$\mathbf{H}=-\partial E/\partial (M \mathbf{m})$, 
where $M$ is the saturation magnetization. 
The energy density is given by 
\begin{equation}
  E 
  =
  -M H_{\rm appl}
  m_{z}
  -
  \frac{M(H_{\rm K}-4\pi M)}{2}
  m_{z}^{2}, 
  \label{eq:energy}
\end{equation}
where $H_{\rm K}$ and $4\pi M$ represent 
the crystalline uniaxial anisotropy and shape anisotropy (demagnetization) fields 
along the $z$-axis, respectively. 
Because the LLG equation conserves the magnetization magnitude, 
the magnetization dynamics can be regarded as the motion of a point particle on a unit sphere, 
as schematically shown in Fig. \ref{fig:fig1}. 
The constant energy curves of Eq. (\ref{eq:energy}) correspond to 
latitude lines on this sphere. 
The spin torque strength $H_{\rm s}$ is \cite{slonczewski96,slonczewski05} 
$H_{\rm s}=\hbar \eta I/[2e (1 + \lambda \mathbf{m}\cdot\mathbf{p})MV]$, 
where $V$ is the volume of the free layer. 
The parameters $\eta$ and $\lambda$ characterize 
the spin torque strength \cite{taniguchi13,slonczewski96,slonczewski05}. 
The third term on the right-hand side of Eq. (\ref{eq:LLG}) represents the random torque 
due to thermal activation, 
where the components of the random field $\mathbf{h}$ satisfy 
the fluctuation-dissipation theorem \cite{brown63} 
$\langle h_{i}(t) h_{j}(t^{\prime}) \rangle = (2D/\gamma^{2}) \delta_{ij} \delta(t-t^{\prime})$, 
where $D=\alpha \gamma k_{\rm B}T/(MV)$ is the diffusion coefficient. 

% ===================================================================================================================================================================================== %

In the self-oscillation state, 
the magnetization precesses almost on a constant energy curve 
because the energy dissipation due to damping 
balances the energy supplied by the spin torque. 
Thus, it is convenient to divide the magnetization dynamics 
into the directions orthogonal to and along the constant energy curve. 
From Eq. (\ref{eq:LLG}), 
the former dynamics is described by
$\dot{E}/(\gamma M)=-\alpha [\mathbf{H}^{2}-(\mathbf{m}\cdot\mathbf{H})^{2}]
  +H_{\rm s}[\mathbf{p}\cdot\mathbf{H}-(\mathbf{m}\cdot\mathbf{p})(\mathbf{m}\cdot\mathbf{H})]
  +(\mathbf{H}\times\mathbf{m})\cdot\mathbf{h}$, 
where the first and second terms on the right-hand side represent 
the energy dissipation due to damping and 
the energy change due to the spin torque, respectively. 
On the other hand, to describe the dynamics along a constant energy curve, 
we introduce the phase $\varphi$ \cite{dykman12}, 
whose dynamics is described by $\dot{\varphi}=(2\pi/\tau)-(2\pi/\tau)[\mathbf{m}\times(\mathbf{m}\times\mathbf{H})]\cdot\mathbf{h}/|\mathbf{m}\times\mathbf{H}|^{2}$. 
Here, $\tau(E)=\oint dt$ is the precession period on the constant energy curve. 
The terms $(\mathbf{H}\times\mathbf{m})\cdot\mathbf{h}$ and $(2\pi/\tau)[\mathbf{m}\times(\mathbf{m}\times\mathbf{H})]\cdot\mathbf{h}/|\mathbf{m}\times\mathbf{H}|^{2}$ 
are the projections of the random torque 
to the $\dot{E}$ and $\dot{\varphi}$ directions, 
and can be replaced with 
$\sqrt{2D [\mathbf{H}^{2}-(\mathbf{m}\cdot\mathbf{H})^{2}]}\xi_{E}/\gamma$ and 
$\sqrt{2D(2\pi/\tau)^{2}/|\mathbf{m}\times\mathbf{H}|^{2}}\xi_{\varphi}/\gamma$ 
with $\langle \xi_{k}(t) \xi_{\ell}(t^{\prime}) \rangle = \delta_{k\ell}\delta(t-t^{\prime})$ ($k,\ell=E,\varphi$), respectively. 
The random torque makes the probability function of the magnetization direction, 
which satisfies the Fokker-Planck equation \cite{dykman12}, 
independent of $\varphi$ \cite{apalkov05,bertotti06} 
after a sufficiently long time compared with $\tau$ passes. 
This, we assume that 
the deterministic torques and diffusion coefficients of 
$\dot{E}$ and $\dot{\varphi}$ can be replaced 
by their averages on the constant energy curve \cite{bertotti06,dykman12}. 
Accordingly, the LLG equation in the energy-phase representation is 
\begin{equation}
\begin{split}
  \frac{d E}{dt}
  =&
  -\frac{M}{\gamma \tau}
  \left(
    \alpha 
    \mathscr{M}_{\alpha}
    -
    \mathscr{M}_{\rm s}
  \right)
  +
  D 
  \left(
    \frac{M}{\gamma}
  \right)^{2}
  \frac{1}{\tau}
  \frac{d \mathscr{M}_{\alpha}}{dE}
\\
  &+
  \frac{M}{\gamma}
  \sqrt{
    \frac{2 D \mathscr{M}_{\alpha}}{\tau}
  }
  \xi_{E}(t), 
  \label{eq:dEdt}
\end{split}
\end{equation}
\begin{equation}
\begin{split}
  \frac{d \varphi}{dt}
  =
  \frac{2\pi}{\tau}
  -
  \sqrt{
    \frac{2D \mathscr{N}_{\varphi}}{\tau}
  }
  \xi_{\varphi}(t),
  \label{eq:dvarphidt}
\end{split}
\end{equation}
where 
$\mathscr{M}_{\alpha}(E)=\gamma^{2} \oint dt [\mathbf{H}^{2}-(\mathbf{m}\cdot\mathbf{H})^{2}]$, 
$\mathscr{M}_{\rm s}(E)=\gamma^{2} \oint dt H_{\rm s} [\mathbf{p}\cdot\mathbf{H} - (\mathbf{m}\cdot\mathbf{p})(\mathbf{m}\cdot\mathbf{H})]$, 
and $\mathscr{N}_{\varphi}(E)=[2\pi/(\gamma\tau)]^{2} \oint dt/|\mathbf{m}\times\mathbf{H}|^{2}$, 
respectively \cite{apalkov05,bertotti06,dykman12,taniguchi13PRB1,taniguchi13PRB2,taniguchi14JAP,comment1}. 
In the first term on the right-hand side of Eq. (\ref{eq:dEdt}), 
the energy dissipation due to damping ($\propto -\alpha \mathscr{M}_{\alpha}$) is always negative, 
whereas the sign of the energy change due to the spin torque ($\propto \mathscr{M}_{\rm s}$) depends on the current direction. 
The second term on the right-hand side of Eq. (\ref{eq:dEdt}) is the thermally generated drift term 
pointed out by Ref. \cite{bertotti06}, 
and represents the energy supplied by the heat bath. 
The last terms of Eqs. (\ref{eq:dEdt}) and (\ref{eq:dvarphidt}) describe 
the fluctuations of the energy and phase, 
where $D_{E}=D(M/\gamma)^{2}\mathscr{M}_{\alpha}/\tau$ and $D_{\varphi}=D \mathscr{N}_{\varphi}/\tau$ 
are the averaged diffusion coefficients of the random torque 
along the $\dot{E}$ and $\dot{\varphi}$ directions, respectively. 
The constant energy curve 
on which the magnetization steadily precesses 
is determined by the condition 
\begin{equation}
  -\frac{M}{\gamma \tau}
  \left(
    \alpha 
    \mathscr{M}_{\alpha}
    -
    \mathscr{M}_{\rm s}
  \right)
  +
  D 
  \left(
    \frac{M}{\gamma}
  \right)^{2}
  \frac{1}{\tau}
  \frac{d \mathscr{M}_{\alpha}}{dE}
  =
  0.
  \label{eq:energy-condition}
\end{equation}

% ===================================================================================================================================================================================== %

Let us calculate the fluctuations of the energy density $E$ and the phase $\varphi$. 
By expanding Eq. (\ref{eq:dEdt}) around the steady-state energy $E$ up to the first order of 
the fluctuation $\delta E$, 
the solution of $\delta E$ is obtained as 
$\delta E(t)=c_{1}e^{-\varOmega_{\alpha}t} + \varXi \int_{-\infty}^{t} d t^{\prime} \xi_{E}(t^{\prime}) e^{-\varOmega_{\alpha}(t-t^{\prime})}$, 
where the term proportional to the integral constant $c_{1}$ is determined by the initial condition, 
and is negligible in the following calculation. 
The quantity $\varOmega_{\alpha}$, 
which characterizes the relaxation of $\delta E$, is defined as 
\begin{equation}
\begin{split}
  \varOmega_{\alpha}
  =&
  \frac{M}{\gamma \tau}
  \frac{d (\alpha \mathscr{M}_{\alpha} - \mathscr{M}_{\rm s})}{dE}
  -
  D  
  \left(
    \frac{M}{\gamma}
  \right)^{2}
  \frac{1}{\tau}
  \frac{d^{2} \mathscr{M}_{\alpha}}{dE^{2}},
  \label{eq:Omega_alpha}
\end{split}
\end{equation}
while $\varXi=(M/\gamma)\sqrt{2D \mathscr{M}_{\alpha}/\tau}$. 
Similarly, the solution of the phase is obtained from Eq. (\ref{eq:dvarphidt}). 
It should be noted that $\tau$ in Eq. (\ref{eq:dvarphidt}) depends on the energy density $E$, 
and therefore, the fluctuation of $E$ should be taken into account 
to determine the phase. 
By expanding $1/\tau$ up to the first order of $\delta E$, 
the solution of the phase is given by 
$\varphi(t)=\varphi(0)+(2\pi t/\tau)-(2\pi/\tau^{2})(d \tau /dE)\int_{0}^{t} dt^{\prime} \delta E(t^{\prime}) - \sqrt{2D \mathscr{N}_{\varphi}/\tau} \int_{0}^{t} dt^{\prime} \xi_{\varphi}(t^{\prime})$, 
where $\varphi(0)$ is the initial phase. 
Then, the phase variance $\Delta\varphi^{2}(t) = \langle \varphi^{2}(t) \rangle - \langle \varphi(t) \rangle^{2}$ is 
\begin{equation}
  \Delta 
  \varphi^{2}(t) 
  =
  2 \Delta \omega_{0}
  \left[
    \left(
      1
      +
      \nu^{2}
    \right)
    |t|
    -
    \nu^{2}
    \left(
      \frac{1-e^{-\varOmega_{\alpha}|t|}}{\varOmega_{\alpha}}
    \right)
  \right],
  \label{eq:phase-variance}
\end{equation}
where $\Delta\omega_{0}=D \mathscr{N}_{\varphi}/\tau$, 
and $\nu^{2}$ is defined as 
\begin{equation}
  \nu^{2}
  =
  \frac{\mathscr{M}_{\alpha}}{\mathscr{N}_{\varphi}}
  \left(
    \frac{2\pi M}{\gamma \tau^{2}\varOmega_{\alpha}}
    \frac{d \tau}{dE}
  \right)^{2}. 
  \label{eq:nu}
\end{equation}
The power spectrum $\mathcal{P}(f)$ of an STO is 
the Fourier transformation of the correlation function 
$\langle \exp\{ i [ \varphi(t) - \varphi(0)] \} \rangle 
  \simeq \exp [ i \langle \varphi(t) - \varphi(0) \rangle ] \exp[-\Delta\varphi^{2}(t)/2]$. 
Then, the linewidth of the power spectrum is determined by the phase variance. 
The power spectrum becomes the Lorentz function 
$\mathcal{P}(f) \propto \Delta f/[(f-f_{0})^{2} + (\Delta f_{\rm L})^{2}]$ 
with the linewidth $\Delta f_{\rm L}$ 
\begin{equation}
  \Delta f_{\rm L}
  =
  \frac{\Delta \omega_{0}}{2\pi}
  \left(
    1
    +
    \nu^{2}
  \right),
  \label{eq:linewidth_Lorentz}
\end{equation}
when $\Delta f_{\rm L} \ll \varOmega_{\alpha}$ \cite{slavin09}. 
In the opposite limit, 
$\mathcal{P}(f)$ is the Gaussian 
$\mathcal{P}(f) \propto \exp[-(f-f_{0})^{2}/(2 \Delta f_{\rm G}^{2})]$, 
with 
\begin{equation}
  \Delta f_{\rm G} 
  = 
  \frac{\sqrt{\Delta \omega_{0}\nu^{2}\varOmega_{\alpha}}}{2\pi}. 
  \label{eq:linewidth_Gauss}
\end{equation}
Equation (\ref{eq:linewidth_Lorentz}) or (\ref{eq:linewidth_Gauss}) is 
applicable to an arbitrary type of STO. 
The term $\Delta\omega_{0}/(2\pi)$ in Eq. (\ref{eq:linewidth_Lorentz}) arises from 
the phase fluctuation 
and is called the phase noise or linear linewidth \cite{slavin09}. 
On the other hand, the term $\Delta\omega_{0}\nu^{2}/(2\pi)$ arises from the energy fluctuation 
and is called the amplitude noise or nonlinear linewidth \cite{slavin09}. 

% ===================================================================================================================================================================================== %

Now let us calculate the linewidth of an STO 
with a perpendicularly magnetized free layer 
and an in-plane magnetized pinned layer. 
Using Eq. (\ref{eq:energy}), 
the energy density $E$ can be directly related to 
the cone angle of the magnetization $\theta=\cos^{-1}m_{z}$. 
The explicit forms of $\tau$, $\mathscr{M}_{\alpha}$, $\mathscr{M}_{\rm s}$, and $\mathscr{N}_{\varphi}$ are 
given by $\tau=2\pi/\{\gamma [H_{\rm appl} + (H_{\rm K}-4\pi M)\cos\theta]\}$, 
$\mathscr{M}_{\alpha}=(2\pi \sin\theta)^{2}/\tau$, 
$\mathscr{M}_{\rm s}=[\pi\gamma\hbar \eta I/(e \lambda MV)](1/\sqrt{1-\lambda^{2}\sin^{2}\theta}-1)\cos\theta$, 
and $\mathscr{N}_{\varphi}=\tau/\sin^{2}\theta$. 
Then, from Eq. (\ref{eq:energy-condition}), 
the relation between the current magnitude $I$ and cone angle $\theta$ of the self-oscillation is given by 
\begin{equation}
\begin{split}
  I
  =&
  \frac{2 \alpha (1-\epsilon) e \lambda MV}{\hbar\eta \cos\theta}
  \left(
    \frac{1}{\sqrt{1-\lambda^{2}\sin^{2}\theta}}
    -
    1
  \right)^{-1}
\\
  &\times
  \left[
    H_{\rm appl}
    +
    \left(
      H_{\rm K}
      -
      4\pi M
    \right)
    \cos\theta
  \right]
  \sin^{2}\theta,
  \label{eq:I_theta}
\end{split}
\end{equation}
where $\epsilon$ is given by 
\begin{equation}
\begin{split}
  \epsilon 
  =&
  \frac{k_{\rm B}T}{M [H_{\rm appl}+(H_{\rm K}-4\pi M)\cos\theta]V \sin^{2}\theta}
\\
  &\times
  \left[
    \frac{2 H_{\rm appl}\cos\theta - (H_{\rm K}-4\pi M)(1-3\cos^{2}\theta)}{H_{\rm appl}+(H_{\rm K}-4\pi M)\cos\theta}
  \right].
  \label{eq:epsilon}
\end{split}
\end{equation}
The term $\alpha(1-\epsilon)$ in Eq. (\ref{eq:I_theta}) represents the fact that 
the magnetization deviates from the $z$-axis 
even if the current is absent 
because of the energy supply $(\propto k_{\rm B}T$) from the heat bath \cite{bertotti06}; 
this effect can be regarded as an effective reduction of the damping constant. 
The deviation angle $\theta_{0}$ from the $z$-axis in the absence of the current 
can be estimated by the condition $1-\epsilon=0$. 
In the zero temperature limit, 
Eq. (\ref{eq:I_theta}) is identical to $I(\theta)$ derived in Ref. \cite{taniguchi13}. 
The threshold current at zero temperature is 
$I_{\rm c}=[4 \alpha e MV/(\hbar \eta \lambda)](H_{\rm appl}+H_{\rm K}-4\pi M)$, 
which can be obtained from Eq. (\ref{eq:I_theta}) 
in the limit of $T \to 0$ and $\theta \to 0$. 
Because the energy dissipation due to the damping balances 
the energy supply from the heat bath 
at the angle $\theta_{0}$, 
an infinitesimal current can excite the magnetization dynamics 
with the cone angle $\theta>\theta_{0}$. 
Thus, our theory gives the linewidth for the current region of $|I|>0$ 
without a separation of the current region at $I_{\rm c}$. 
This differs from the previous theory \cite{slavin09}, 
in which the magnetization is fixed in the equilibrium direction for $I<I_{\rm c}$, 
so the linewidth due to thermal activation becomes finite only for $I>I_{\rm c}$. 
The explicit forms of Eqs. (\ref{eq:linewidth_Lorentz}) and (\ref{eq:linewidth_Gauss}) are 
\begin{equation}
  \Delta f_{\rm L}
  =
  \frac{\alpha\gamma k_{\rm B}T}{2\pi MV \sin^{2}\theta}
  \left[
    1
    +
    \left(
      \frac{\omega_{\rm K}}{\varOmega_{\alpha}}
    \right)^{2}
    \sin^{4}\theta
  \right],
  \label{eq:linewidth_1}
\end{equation}
and $\Delta f_{\rm G}=[\omega_{\rm K}/(2\pi)]\sqrt{\alpha\gamma k_{\rm B}T/(MV\varOmega_{\alpha})}\sin\theta$, 
where $\omega_{\rm K}=\gamma(H_{\rm K}-4\pi M)$. 
In $\Delta f_{\rm G}$, 
$1 \ll \nu^{2}=(\omega_{\rm K}/\varOmega_{\alpha})^{2}\sin^{4}\theta$ is again assumed. 
The explicit form of $\varOmega_{\alpha}$ is 
\begin{equation}
\begin{split}
  \varOmega_{\alpha}
  =&
  \alpha
  \gamma
  \left[
    2 H_{\rm appl}
    \cos\theta
    -
    \left(
      H_{\rm K}
      -
      4\pi M
    \right)
    \left(
      1
      -
      3 \cos^{2}\theta
    \right)
  \right]
\\
  &+
  \frac{\gamma \hbar \eta I}{2e \lambda MV}
  \left[
    \frac{1- \lambda^{2} - (1-\lambda^{2}\sin^{2}\theta)^{3/2}}{(1-\lambda^{2}\sin^{2}\theta)^{3/2}}
  \right]
\\
  &+
  D 
  \left\{
    3
    -
    \frac{H_{\rm appl}^{2}-(H_{\rm K}-4\pi M)^{2}}{[H_{\rm appl}+(H_{\rm K}-4\pi M)\cos\theta]^{2}}
  \right\}.
  \label{eq:varOmega_1}
\end{split}
\end{equation}
These equations give 
the relation between the current, cone angle of the oscillation, and linewidth. 

% ===================================================================================================================================================================================== %

\begin{figure}%[p]
\centerline{\includegraphics[width=1.0\columnwidth]{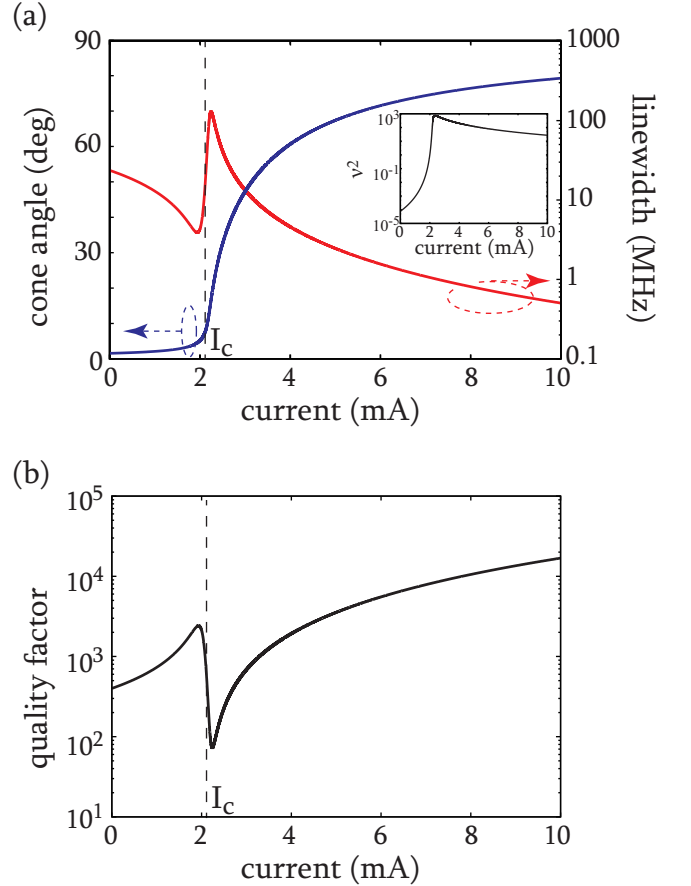}}\vspace{-3.0ex}
\caption{
         (a) Dependence of the cone angle $\theta$ 
             and linewidth $\Delta f_{\rm L}$ (Eq. (\ref{eq:linewidth_1})) on the current magnitude. 
             The dashed line indicates the threshold current at zero temperature, $I_{\rm c}=2.2$ mA. 
             Inset shows the current dependence of $\nu^{2}$. 
         (b) Dependence of the quality factor on the current magnitude. 
         \vspace{-3ex}}
\label{fig:fig2}
\end{figure}

% ===================================================================================================================================================================================== %

% ===================================================================================================================================================================================== %

Figure \ref{fig:fig2} (a) shows 
the dependence of the cone angle $\theta$ and the linewidth $\Delta f_{\rm L}$ 
on the current, 
where the values of the parameters are 
$M=1448$ emu/c.c., 
$H_{\rm K}=18.6$ kOe, 
$\gamma=1.732 \times 10^{7}$ 1/(Oe$\cdot$s), 
$V=\pi \times 60 \times 60 \times 2$ nm${}^{3}$, 
$\alpha=0.005$, 
$\eta=0.54$, 
$\lambda=\eta^{2}$, 
$H_{\rm appl}=3.0$ kOe
and $T=300$ K \cite{kubota13}. 
Using these values, 
$\varOmega_{\alpha}$ is typically on the order of $0.1-1$ GHz, 
satisfying the condition that $\mathcal{P}$ to be the Lorentz function, $\Delta f_{\rm L} \ll \varOmega_{\alpha}$. 
Therefore, the linewidth of the Lorentz function is shown in Fig. \ref{fig:fig2} (a). 
The deviation angle of the magnetization from the $z$-axis, 
which satisfies $1-\epsilon=0$, is $\theta_{0}=1.6$ deg. 
The threshold current at zero temperature is $I_{\rm c}=2.2$ mA. 
From $\theta_{0}$, the cone angle $\theta$ monotonically increases 
as the current increases. 
On the other hand, the linewidth above $I_{\rm c}$ 
monotonically decreases as the current increases, 
whereas that below $I_{\rm c}$ depends nonmonotonically on the current. 
Such complex current dependence of the linewidth 
was found in previous experimental studies \cite{kubota13,mistral06}. 
Above $I_{\rm c}$, 
the length of the precession trajectory becomes long 
as the current, as well as the cone angle, increases. 
Then, the phase fluctuation by one random torque becomes relatively small as the current increases, 
resulting in linewidth decreases. 
On the other hand, below $I_{\rm c}$, 
not only the fluctuation but also the effective reduction of the damping constant, $\alpha(1-\epsilon)$, 
by the energy supplied by the heat bath affects the linewidth. 
Whereas the linewidth due to the former contribution increases as the current decreases, 
the linewidth due to the latter contribution decreases. 
Because of this competition, 
the linewidth shows a nonmonotonic dependence on the current below $I_{\rm c}$. 
The current dependence of $\nu^{2}$ is shown in the inset of Fig. \ref{fig:fig2}, 
indicating that the amplitude noise dominates the linewidth 
in the current region of $I > I_{\rm c}$. 

% ===================================================================================================================================================================================== %

Figure \ref{fig:fig2} (b) shows the current dependence of the Q-factor $f_{0}/\Delta f_{\rm L}$, 
where $f_{0}=1/\tau$. 
The Q-factor reaches $10^{4}$ for the current $I \simeq 10$ mA, 
which is two orders of magnitude larger than the experimentally observed value (135 in Ref. \cite{kubota13}). 
This result should motivate future experimental research for practical STO applications. 
The current magnitude $I \simeq 10$ mA is larger than 
the experimentally available applied current (typically, $3$ mA \cite{kubota13}) 
for an MgO-based magnetic tunnel junction. 
A reduction in the resistance due to the tunnel barrier will be necessary 
to apply a large current. 
A giant magnetoresistive system would also be an interesting candidate 
for this purpose. 

% ===================================================================================================================================================================================== %

The linewidth shown in Fig. \ref{fig:fig2} (a) 
is narrower than the experimental value. 
For example, the calculated linewidth in Fig. \ref{fig:fig2} (a) is, at maximum, about 100 MHz. 
On the other hand, 
the experimental value is, at minimum, 50 MHz, 
and is typically much wider than 100 MHz \cite{kubota13}. 
This fact indicates that other physical mechanisms contribute to 
the linewidth in the experiments. 
One possible contribution is the shot noise effect, 
which arises from the discrete transmission of electrons through the tunnel barrier 
and produces fluctuations of the current, as well as the spin torque. 
This fluctuating spin torque can be effectively included in the random torque 
by renormalizing the damping constant $\alpha$, 
and leads to an increase in the linewidth \cite{dykman12}. 
Another possibility is 
a reduction in the perpendicular anisotropy observed in the experiment \cite{kubota13}, 
whose origin was assumed to be Joule heating in Ref. \cite{kubota13}. 
The reduction in $H_{\rm K}$ make the fluctuations of $E$ and $\varphi$ relatively large, 
which leads to an increase in the linewidth. 

% ===================================================================================================================================================================================== %

Finally, let us discuss the relation between 
the present work and previous theoretical work. 
Slavin and Tiberkevich \cite{slavin09} calculated 
the linewidth by a similar approach, 
but using the oscillation amplitude of the magnetization 
($p$ in their notation) 
instead of the energy density $E$. 
In Ref. \cite{slavin09}, 
the magnetization direction of the pinned layer is assumed to be 
parallel to the easy axis of the free layer, 
and the spin torque parameter $\lambda$ is neglected. 
However, a finite $\lambda$ should be taken into account 
to define physical quantities such as the threshold current $I_{\rm c}$ \cite{taniguchi13} and linewidth 
when the pinned layer magnetization is orthogonal to 
the easy axis of the free layer, 
as is the case for the STO studied in this letter. 
More generally, when the constant energy curve is symmetric with respect to an axis 
perpendicular to the pinned layer magnetization, 
$\lambda$ should be finite. 
In fact, if $\lambda$ is set to zero, 
$I_{\rm c} \to \infty$, 
and the linewidth becomes independent of the current, 
which clearly contradicts the experiments. 
Mathematically, Ref. \cite{slavin09} assumed that 
$\mathbf{p}$ is parallel to the easy axis of the free layer, 
and $H_{\rm s}$ is independent of $\mathbf{m}$ and $\mathbf{p}$ 
in the calculation of $\mathscr{M}_{\rm s}$. 
On the other hand, we did not make such an assumption 
when calculating $\mathscr{M}_{\rm s}$. 
The energy supplied by the heat bath \cite{bertotti06}, 
which is expressed by the second term of Eq. (\ref{eq:dEdt}), 
is also neglected in Ref. \cite{slavin09}. 
In light of these differences, our formula can be regarded as 
a generalization of the formulae in Ref. \cite{slavin09}. 

% ===================================================================================================================================================================================== %

In conclusion, 
we developed a general formula for the linewidth of an STO  
by solving the LLG equation in the energy-phase representation. 
We applied the formula to an STO 
with a perpendicularly magnetized free layer 
and an in-plane magnetized pinned layer 
to estimate the linewidth and quality factor. 
It was shown that the linewidth is suppressed to 0.1 MHz 
when a large current is applied. 
A quality factor larger than $10^{4}$ is predicted in the large current limit, 
which is two orders of magnitude larger than the value recently observed in experiments.

% ===================================================================================================================================================================================== %

%\begin{figure}[p]
%\centerline{\includegraphics[width=1.0\columnwidth]{fig1.eps}}\vspace{-3.0ex}
%\caption{
%         \vspace{-3ex}}
%\label{fig:fig1}
%\end{figure}

% ===================================================================================================================================================================================== %

The author acknowledges H. Imamura, T. Yorozu, H. Maehara, S. Tsunegi, H. Tomita, and H. Kubota. 
This work was supported by JSPS KAKENHI Grant-in-Aid for Young Scientists (B) 25790044. 

% ===================================================================================================================================================================================== %

%\bibliography{biblist}% Produces the bibliography via BibTeX.

\begin{thebibliography}{26}
\bibitem[{com()}]{comment1}
\bibinfo{note}{$\mathscr{M}_{\alpha}$ and $\mathscr{M}_{\rm s}$ of the first
  term in Eq. (\ref{eq:dEdt}), and $D \mathscr{M}_{\alpha}$ and $D
  \mathscr{N}_{\varphi}$ in the random torques of Eqs. (\ref{eq:dEdt}) and
  (\ref{eq:dvarphidt}) are proportional to $\int_{0}^{2\pi} (d\varphi/2\pi)
  F(E,\varphi)$, $\int_{0}^{2\pi} (d\varphi/2\pi)
  \mathbf{I}_{s}\cdot\mathbf{V}(E,\varphi)$, $\mathcal{D}_{E}$, and
  $\Delta_{\bar{E}}$ of Ref. \cite{dykman12}, respectively.}
\end{thebibliography}

% ===================================================================================================================================================================================== %

\end{document}